\documentclass[a4paper,fleqn,usenatbib]{mnras}

\usepackage{newtxtext,newtxmath}

\usepackage[T1]{fontenc}
\usepackage{ae,aecompl}

\usepackage{graphicx}	
\usepackage{amsmath}	
\usepackage{amssymb}	

\title[Short Periodic variability of Mrk 231]{Confirmed short periodic variability  of subparsec supermassive binary black hole  candidate Mrk 231}

\author[A. B. Kova{\v c}evi{\'c }et al.]{
Andjelka  B. Kova{\v c}evi{\' c},$^{1}$\thanks{E-mail: andjelka@matf.bg.ac.rs (ABK)}
Tignfeng Yi,$^{2}$
Xinyu  Dai,$^{3}$
Xing Yang,$^{2}$
Iva {\v C}vorovi{\'c}-Hajdinjak$^{1}$
 \newauthor {and Luka {\v C}. Popovi{\' c}$^{1,4}$  }
\\
$^{1}$Department of astronomy, Faculty of Mathematics, University of Belgrade, Studentski trg 16, 11000, Belgrade, Serbia \\
$^{2}$Key Laboratory of Colleges and Universities in Yunnan Province
for
High-energy Astrophysics, Department of Physics, \\
Yunnan Normal University, Kunming
650500, China\\
$^{3}$Homer L. Dodge Department of Physics and Astronomy, University of Oklahoma, Norman, OK 73019, USA\\
$^{4}$ Astronomical observatory Belgrade, 
Volgina 7, P.O.Box 74 11060, Belgrade,  11060, Serbia
}

\date{Accepted XXX. Received YYY; in original form ZZZ}

\pubyear{}

\begin{document}
\label{firstpage}
\pagerange{\pageref{firstpage}--\pageref{lastpage}}
\maketitle

\begin{abstract}

Here we confirm the short periodic variability of a subparsec supermassive binary black hole (SMBBH)   candidate Mrk 231  in  the  extended optical photometric data set collected by    the Catalina Real-Time Transient Survey (CRTS) and All-Sky Automated Survey for Supernovae (ASAS-SN). 
Using the Lomb-Scargle periodogram and 2DHybrid method, we detected the significant  periodicity   of $\sim 1.1$ yr   beyond a damped random walk model in the CRTS$+$ASAS-SN optical data set.
Mrk 231 has been previously proposed as a SMBBH candidate with a highly unequal mass ratio ($q\sim 0.03$),  very tight mutual separation of $\sim 590$ AU, and an orbital period of  $\sim 1.2$ yr.
Hence, our result further supports, even though  not prove, the intriguing hypothesis that
SMBBHs with low mass ratios may be more common than close-equal mass SMBBHs.
This result, however, was  obtained  from  the contribution of  CRTS data  with limited sampling cadence and  photometric accuracy, and   further monitoring of  Mrk 231 is crucial to confirm the periodicity.

\end{abstract}




\section{Introduction}

Close, gravitationally bound supermassive binary black holes  (SMBBHs) are theoretically predicted a long time ago as a consequence   of the galaxy  mergers \citep{1980Natur.287..307B}.
As  galaxy mergers should be frequent 
with the expectation that SMBBHs linger at very close mutual distance \citep{ 2002MNRAS.336L..61H}, it is believed that sub-parsec  SMBBHs should be  common, despite the lack of observational conformations. Now, these intriguing objects are the prominent targets  for  imminent   gravitational wave surveys. 
Recently, \cite{2019ApJ...879L..G} reported the discovery  of  a $z \sim 0.2$ quasar residing in a merger remnant with two closely separated ($\sim 430 $ pc)
continuum cores at the center of the galaxy SDSSJ1010$+$1413.   These two cores are spatially matched  with two
powerful $[$OIII$]$-emitting point sources with  luminosities 
$\sim 5\times 10^{46} \mathrm{{erg}{s}}^{-1}$,
 indicating  a bound  SMBBH system (each object in the  pair with $ 4\times 10^{8}M{\odot}$). This finding gives strong support that detection of such systems is possible in the gravitational  wave domain.
Still, most SMBBHs  angular sizes are well below the resolution of direct imaging methods and some  other indirect methods are used to  detect
sub-pc SMBBHs presumably from spectroscopic line shapes and time variability of long term monitored continuum and emission lines \citep[see e. g.][]{10.1038/nature07779, 2012ApJ...759..118B,2015MNRAS.453.1562G, 10.1038/nature14143, 2016ApJ...822....4L}.
Yet as more measurements and analyses have accumulated, that spectral line shapes can not be uniquely coupled to a spatial model of the SMBBHs,  the line shape method has seemed increasingly inadequate to use alone \citep[see simulations in][]{10.1093/mnras/stz1713,2019arXiv191008709K}.

With the development of  time domain astronomy, the identification of SMBHBs via periodic variability of light curves hold some promises \citep[see e.g.][]{2009ApJ...700.1952H, 2018ApJ...856...42S}. 
This working  hypothesis is based on the theoretical prediction that if gas accretion occurs in  SMBBH, the orbital period could be imprinted into periodic variability of the continuum and emission lines.

Proposed spectroscopic SMBBH signatures  (e.g. double-peaked
profiles)  have been identified  in large galaxy monitoring
samples \citep{2012ApJS..201...23E,2015MNRAS.453.1562G}. Even though the results are encouraging, still there are  alternate physical explanations for such emission \citep[see e. g.][]{2012NewAR..56...74P}.

Also,  the statistical significance of the detected periodic signals is strongly influenced by the   underlying stochastic quasar
 variability and quality of data time coverage.  \cite{2016MNRAS.461.3145V} have
recently shown that Gaussian red noise models can naturally
mimic periodic signals, especially
at inferred periods comparable to the length of the observed light curves. 
Moreover, optical surveys  are expected to detect
SMBBHs  at  larger periods  where they stay for 
the largest fraction of their lifetimes \citep[see further discussion in ][]{2018ApJ...856...42S}. 

Thus, most binary SMBBH candidates require confirmation with newly available data and newly devised methods for period detection. Upcoming synoptic instruments like LSST 
will potentially identify hundreds to thousands of such candidates  \citep{https://agn.science.lsst.org/sites/default/files/LSST_AGN_SC_Roadmap_v1p0.pdf}, thus emphasizing the need for new, and effective methods for period detection. From the other side, it is expected  that the Event Horizon Telescope  \citep{10.3847/2041-8213/ab1141}  can spatially resolve sub-pc SMBBH, and such confirmations will  not be  limited by the  periodicity detection method or  by a condition that  source orbital period should be much larger than  the monitoring program. 

Recently, \cite{2015ApJ...809..117Y}  proposed  an interesting SMBBH candidate (with a very short orbital period) in the nucleus of the nearest quasar Mrk 231 at redshift $ z = 0.0422$, based on detailed analysis of  its unique optical-UV spectrum. 
Mrk 231 is the nearest quasar  and the most luminous Ultra-Luminous InfraRed Galaxy in the nearby Universe \citep{10.1051/0004-6361/201015164}.
Lately, the emission feature in its spectra has been reported as the  first detection of extragalactic molecular oxygen \citep{10.3847/1538-4357/ab612d}.
Even the optical  spectrum of Mrk 231  resembles mostly  the
quasar composite spectrum,  it varies  dramatically at
the wavelengths around 3000 \AA\    and becomes flat again at
$< 2500$ \AA. They interpreted  this unique  spectrum of Mrk 231 as a consequence of  emission from a BBH accretion system, with
which the drop of the continuum at $< 4000$ \AA\  is due to a gap or
a hole opened by the secondary component of the BBH \citep{ 10.1088/0004-637X/761/2/90}.
The masses of the primary and
the secondary SMBBHs were constrained as $\sim  1.5 \times 10^8$
M$\odot$ and $4.5\times 10^ 6$ M$\odot$,
respectively.
The semimajor
axis of the system  is estimated  $\sim 590$ AU, and its orbital
period is just $\sim 1.2$ years, where the  corresponding  gravitational wave  emission would be on tens of
nanohertz. If such  periodicity could be confirmed from timing observations, 
this object might be a target for the next generation of  gravitational wave detections.

The existence of short time scale periodic signals  have been reported in the
optical domain for some other active galactic nuclei (AGN). \cite{2014ApJ...793L...1S, 2016AJ....151...54S}
 confirmed  315 day period in the light curve of
PKS 2155-304 detected previously by \cite{ https://iopscience.iop.org/article/10.1088/1674-4527/14/8/004/meta},
and also  found marginally
significant ($3\sigma$) periods in PKS 0537-441, OJ 287,
3C 279, PKS 1510-089 and PKS 2005-489, on timescales
ranging from dozens of days up to a few years (some of them in
harmonic relations).
Also, \cite{2016AJ....151...54S}
found a marginally-significant ($3\sigma$) signal in the optical
light curve of OJ 287 with a period of 435 days,
in addition to a much less significant one of 203 days.

Having in mind a prior model prediction of a short  periodic signal in Mrk  231  and SMBBH  model evaluation is an iterative process, we  analyze  the combined CRTS and ASAS-SN light curves of Mrk 231.
Our analysis   and extracted results are obtained from   for the first time presented   13 year long
 optical light
curve of Mrk 231, utilizing the traditional Lomb-Scargle Periodogram,
and the newly introduced 2DHybrid method. We estimate the statistical significance  of the identified signal by generating synthetic light curves that show red noise without any periodic signal.
This paper is organized as follows.  We present, for the first time, combined photometric Mrk 231 light curve  in Section \ref{obs}   and used in Section  \ref{perdet} for periodicity analysis.  The main results are summarized in Section \ref{sum}.


\section{Observations}{\label{obs}}

The data sets, we used for the analysis, consist of the   V-band light curves obtained by CRTS and ASAS-SN surveys.
The data of the  V-band light curve of Mrk 231 was taken from  The Catalina Surveys Data Release 2 (CSDR2) \citep{2009ApJ...696..870D,2013ApJ...763...32D, 2015MNRAS.453.1562G}.
The light curve of  the CSRT DR2 V-band  contains 113 epochs data collected between May 2005 and June 2013. The effective wavelength of the V-band is 5510 \AA.

The ASAS-SN V-band data were recorded  by the Brutus (Haleakala, Hawaii) and
Cassius (CTIO, Chile) quadruple telescopes between 2013 and 2018. ASAS-SN
V-band field is observed to a depth of $V\lesssim17$ mag. The field of view of an
ASAS-SN camera is 4.5 deg$^2$, the pixel scale is 8\farcs0 and the FWHM is
typically ${\sim}2$ pixels. The ASAS-SN V-band light curves were processed via method explained in \citet{2018MNRAS.477.3145J} using image subtraction
\citep{1998ApJ...503..325A,2000A&AS..144..363A} and aperture photometry on the
subtracted images with a 2-pixel radius aperture. The APASS catalog
\citep{2015AAS...22533616H} was used for calibration. We corrected the zero point
offsets between the different cameras as described in \citet{2018MNRAS.477.3145J}.
The photometric errors were estimated as prescribed by
\citet{2019MNRAS.485..961J}.

\begin{table}
	\centering
	\caption{The combined V band light curve of Mrk 231.  The columns display Modified Julian Date, magnitude and magnitude uncertainties, respectively. Complete data are  available online.  }
	\label{tab:celo}
	\begin{tabular}{ccc} 
		\hline
MJD& mag&$\sigma$\\
\hline
53505.21965  &  13.034  &  0.06  \\
53767.31269  &  13.214  &  0.05  \\
53767.31765  &  13.204  &  0.05  \\
53767.32261  &  13.214  &  0.05  \\
53767.32757  &  13.214  &  0.05  \\
53860.36368  &  13.244  &  0.05  \\
53860.37032  &  13.264  &  0.05  \\
53860.37695  &  13.254  &  0.05  \\
53860.38362  &  13.244  &  0.05  \\
53884.24010  &  13.274  &  0.05  \\
53767.32261  &  13.214  &  0.05  \\
  ...                   &  ...         &  ...\\
\hline
	\end{tabular}
\end{table}

\begin{figure}
	\includegraphics[width=\columnwidth]{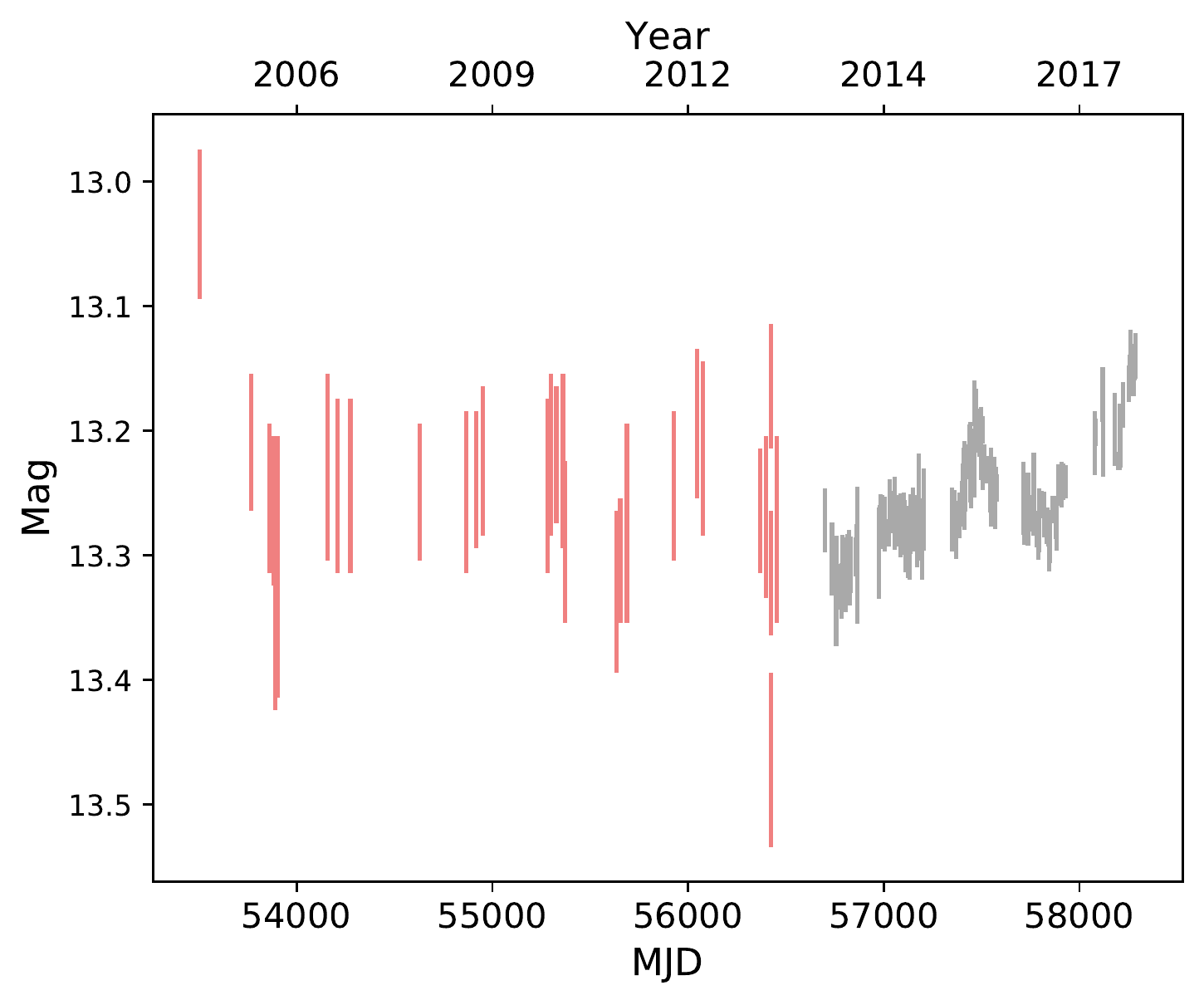}
    \caption{The combined long-term   V band light curve of Mrk 231. This consists of  CRTS  (red error bars, {\boldmath MJD$<$56500}) and ASAS-SN (black errorbars, {\boldmath MJD$>$56500})  data points. To create the combined record we use magnitude difference $\delta m=0.304$.
    The errorbars stand for $1 \sigma$ standard deviation of the measurement. Time is given as Modified Julian days (MJD).}
    \label{fig:curve}
\end{figure}

 However, due to   differences in the wavelength response of telescopes, and zero- point calibration systematic offsets between light curves from CRTS and ASAS-SN telescopes can be present.  We tested for these offsets by calculating the differences  between the  average values of the two light curves. 
The  mean magnitude of CRTS and ASAS-SN  data are 12.952 and 13.256 respectively, so
the difference between them is 0.304. We let the CRTS data to be translated for  0.304 magnitudes  to be
consistent with  the ASAS-SN data (see Fig. \ref{fig:curve}). 
We applied these calculated shifts of the mean values of two light curves to produce the combined light curve (see Table \ref{tab:celo}). Table \ref{tab:celo}  is available in the electronic supplementary material.

In Table \ref{tab:prva} we show some characteristic parameters of the combined CRTS and ASAS-SN light curves. This shows that the ASAS-SN part of the light curve is more variable than the  CRTS part, thus contributing more to the variability of  the combined curve. CRTS part of the light curve is almost three times less variable than the ASAS-SN part. Moreover, the ASAS-SN mean sampling rate is almost three times better than CRTS. Also, CRTS data is more sparse with larger
season gaps, and ASAS-SN data is denser with smaller time gaps.
As a summary,  these characteristics might negatively influence the quality of results based on the  CRTS data set.

\section{Methods of periodicity detection in the light curves}{\label{perdet}}

There are many  characteristics of the astronomical observations, which  severely
complicate period analyses: the limited length of the light curves, the presence of gaps in the light curves (e.g.,
seasonal observing opportunities, seeing, etc.) and uneven sampling as well. Extended
gaps in time series are difficult because their duration can be
comparable to the duration of the observations in a year. This problem is evident in the optical  light curve
of Mrk 231 (Fig. \ref{fig:curve}). For this purpose,  we applied two different procedures for periodicity detection which are  briefly  outlined below.

\subsection{Lomb-Scargle periodogram}{\label{metlsp}}

An extension of the Fourier technique
that attempts to circumvent  above mentioned  limitations is the
Lomb-Scargle periodogram \citep[LSP][]{1976Ap&SS..39..447L, 1982ApJ...263..835S} .
The LSP  is the most widely used statistical tool  for  detecting   periodic signals in unevenly-spaced observations in the last few  decades.
It utilizes  correction of  the functional
basis  of the Fourier transform to
sustain  normalization condition on an uneven grid of
time instances  of observation. The prominent attributes of
the Lomb-Scargle normalized periodogram are avoidance of  the  interpolation  of missing data and a  per-point rather than a per-time-interval  weighting observations  \citep{10.1086/304206}.

We used the LSP implementation in astropy package of Python. For a detailed practical discussion of LSP  based on astropy see \cite{2018ApJS..236...16V}. 
Some odd characteristics of LSP has been noted. The LSP can be sensitive to frequencies higher than the average Nyquist frequency. The periodogram can have many spurious peaks, which can arise due to several factors.
For example, when the  signal is not a perfect sinusoid, additional peaks can appear indicating higher-frequency components in the signal.  Uncertainties  in observations could cause loss of power from the true peaks.
Also, the LSP method is based on a  least square regression and, its results can be   reactive  to outliers in the light curve.
The consequence of these effects means that the highest LSP peak  may not correspond to the most probable  frequency, and the analysis  results must be interpreted carefully \citep[see discussion in][]{2018ApJS..236...16V}.
\begin{figure}
	\includegraphics[width=\columnwidth]{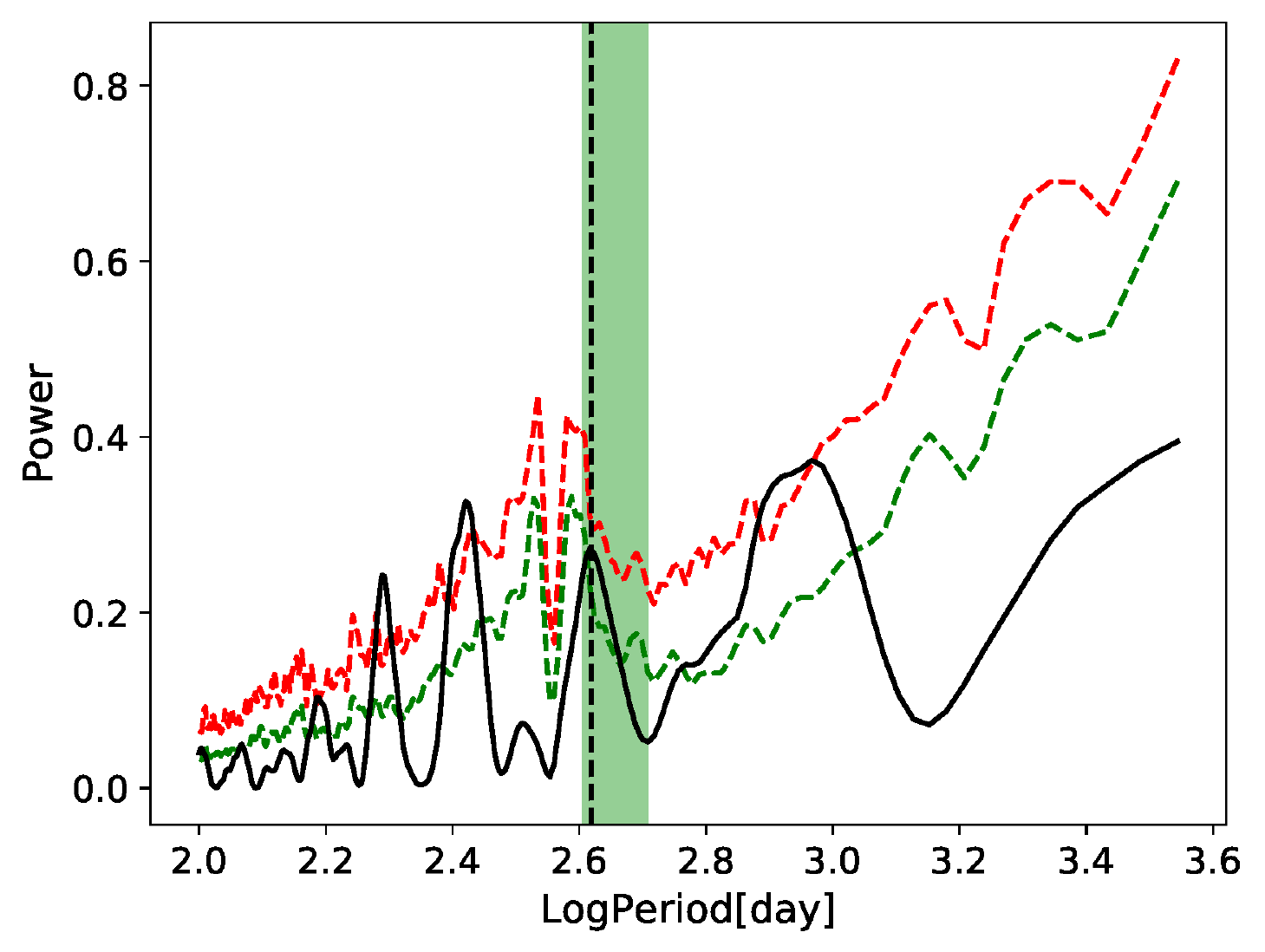}
    \caption{The LSP  computed from  the photometric light curve of Mrk231 (black) shown in Fig. \ref{fig:curve}, 
and the corresponding 95$\%$ and 99.7$\%$ confidence curves (green (lower) and red (upper) dashed line), for a red noise spectrum with $\alpha=1.85$ (see Subsection \ref{redef}). The vertical dashed line denotes the period of 416 days. The green shaded region indicates a range  for Yan et al (2015) periodicity prediction,
which is  $1.2^{+0.2}_{-0.1} $yr. }
    \label{fig:LSP}
\end{figure}

The calculated
LSP of the optical  light curve in the V   band is presented
in Fig \ref{fig:LSP} and in Table \ref{tab:prva}, which clearly shows 3 peaks of similar amplitude and width.
These three peaks are very close to each other: 195 days, 265 days and 416 days (1.13 yr).
The difference between the first two peaks is 70 days (or about two months), which is not related to the sampling patterns of the curves (see Table \ref{tab:prva}).
The  uncertainties are obtained as the Full-widths
at half-maxima (FWHM) of Gaussian fits to the corresponding
periodogram peaks.

\subsection{2DHybrid method}{\label{methyb}}

Although the LSP method accounts for irregular spacing
in a time series, the method assumes sinusoidal forms of the underlying signal and does  not take into account
time fluctuations in the periodic signal information.
In real astronomical systems,  oscillations
can be damped or deformed. In such cases, the wavelet transform
method proves to be a more useful tool, and it
is often applied in the analysis of blazar sources \citep[see e. g.][]
{2013A&A...558A..92B, 2008A&A...488..897H}.

The innovative algorithm for extraction of periodic signals, 2DHybrid, can robustly distinguish between pure stochastic red noise and a mixture of red noise plus periodic signal in AGN light curves \citep{ 2018MNRAS.475.2051K,2019ApJ...871...32K}. 
In the interest  of completeness, we briefly summarize the method here. Simply knowing the continuous wavelet transform matrix ($ \mathcal{T}_i$) of the light curve 
($\mathrm{c}_{i}$), we estimate the envelope of present  signals 
$ \mathcal{E}_{i}=\sqrt{\mathcal{T}_{i}\cdot\mathcal{T}_i}$, 
where operations are on complex matrices. Then we correlate the envelopes of the light curves ($\mathrm{c}_{i}$, $\mathrm{c}_{j}$)  as follows
$\mathcal{R}=\mathrm{corr}(\mathcal{E}_{i},\mathcal{E}_{j})$. In the case of Mrk 231, we have only one curve, and all formulas are used  for $i=j$. 
The 2DHybrid method is general and different from other methods (such as LSP) since it does not assume any form of signal and its principles are similar to 2D spectroscopy. Instead of the typical 1D representation of frequencies, 2DHybrid method extends the search for signals in 2D, quantifying the strength of the interaction between oscillations in different light curves as well as the level of the autocorrelation of signals in a single curve, which is important in classification of different physical origin of AGN variability. For example, besides periodic oscillations typical e. g. SMBBH candidates, the 2DHybrid allows classification of other types of objects according to fluctuations of oscillations in 2D (auto)correlation space.

\begin{figure}
	\includegraphics[width=\columnwidth]{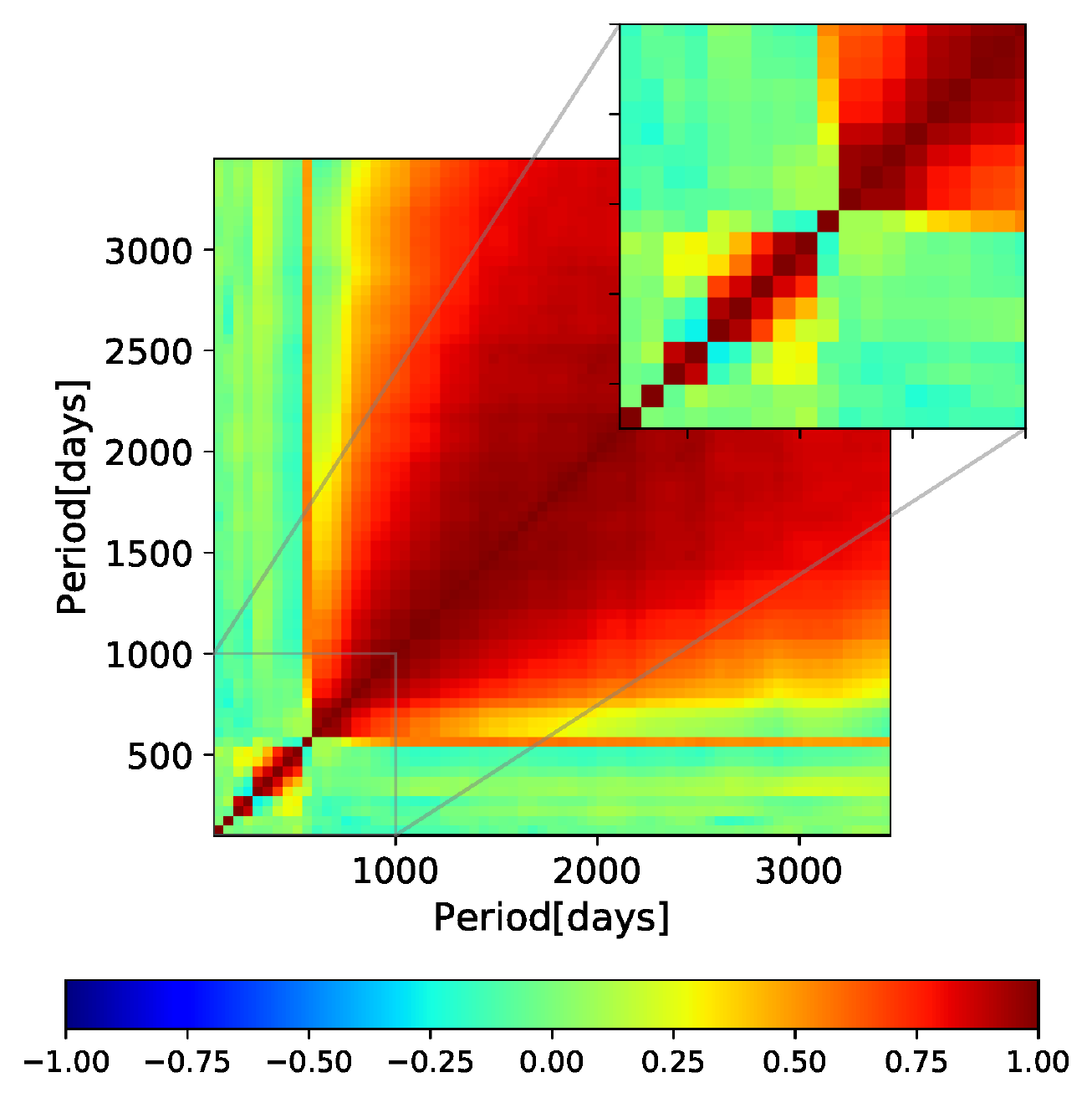}\\	
	\includegraphics[width=\columnwidth]{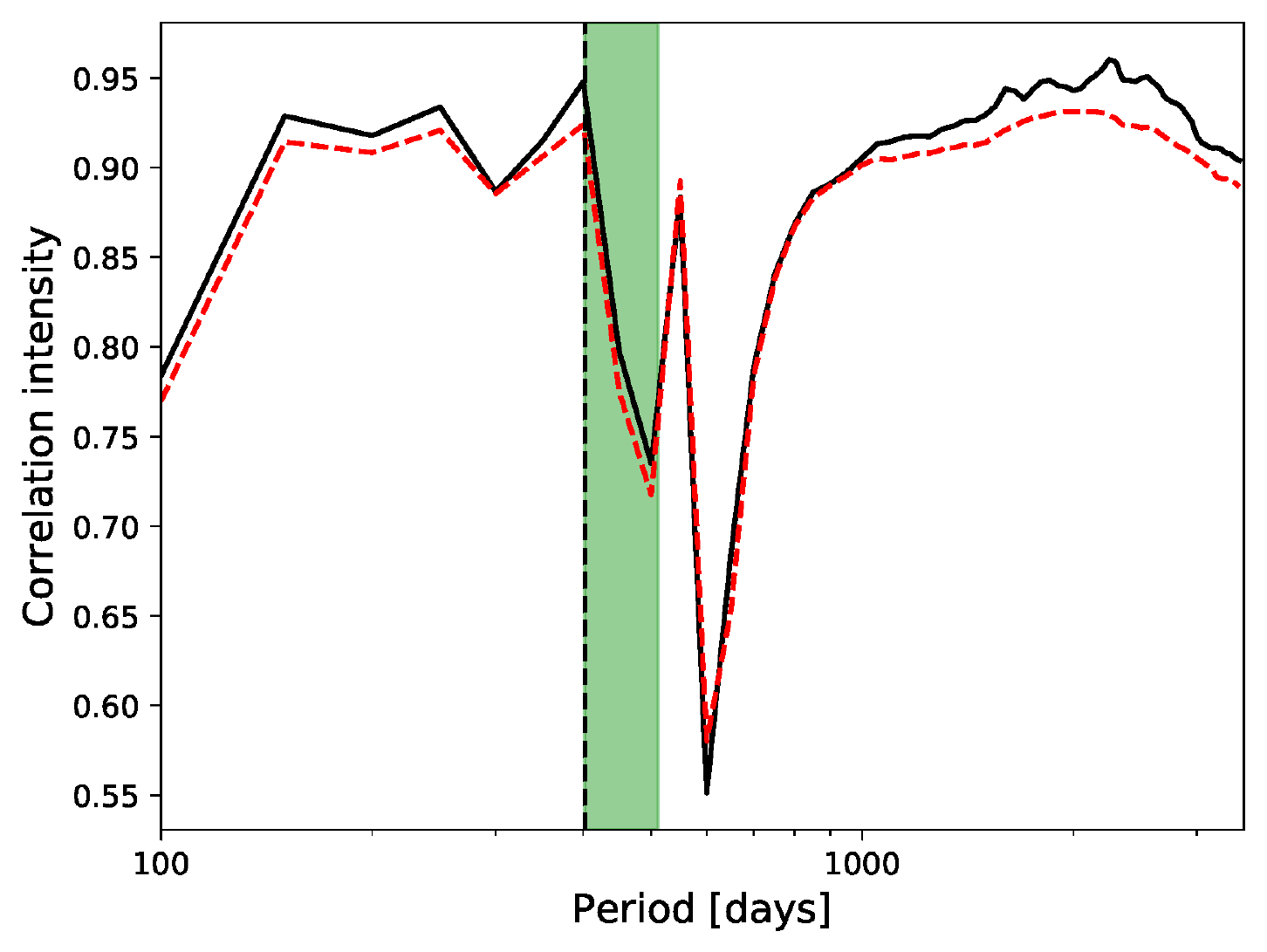}
    \caption{Upper panel: The color-scaled correlation intensity (red for high correlation intensities) of  coupling of periods in the combined light curve of  Mrk 231 in the period-period plane. Three prominent periods  are present in red in the lower-left corner ($\sim$195,$\sim$265 and $\sim$403 days, from left to the right).  Correlation islands ($\sim$0.25) indicate  an asynchronous coupling between different periods.  
Inset: zoomed region in period-period  plane between 100 and 1000 days, where three periods are  seen.     
    Lower panel: Time averaged 2DHybrid
correlation  intensity as a function of the period (solid black curve),  and $99.7 \%$ confidence
curve (red dashed-line) from
the simulation. The vertical dashed line marks the period of 403 days (1.1 yr). The green shaded region indicates a range  for Yan et al (2015) periodicity prediction,
which is  $1.2^{+0.2}_{-0.1} $yr. }
    \label{fig:hyb}
\end{figure}

The  2DHybrid method results  (see Figure  \ref{fig:hyb} and Table \ref{tab:prva}) indicate
that the correlation intensity  for the characteristic periods centered at
403 days  is highly significant (larger than 99.7$\%$), which is similar to  the  LSP result of period 413 days with significance larger than 95$\%$.  Their uncertainties are evaluated as
the means of the FWHMs of the Gaussian fits centered
around the peaks at a given period.  The relatively larger uncertainty in the 2DHybrid-derived periods mainly arises from the  temporal variation of the periodicities.
The average oscillation power is quite similar to the LSP power.

\subsection{Estimation of red noise effects}\label{redef}

As already mentioned, the effects of uneven sampling of a light
curve   can  produce spurious peaks in
the periodogram that can be mistaken for a real signal. Moreover, 
the LSP and 2DHybrid methods in Mrk 231 light curve, we take into account both the uneven
sampling including seasons gaps of the object light curve and the red noise.
To do so, we 
simulate one thousand   red-noise light curves
of the source, obtained  based on 
\cite{1995A&A...300..707T} method.

First, as an input parameter  for simulating the  red noise light curves, we  calculate   the power law index of the
underlying  red noise of the source. We used   a broken power-law of the form
$P(\nu) \sim [1 + (\nu/\nu_{b})^2]^{-\alpha}$ where $\alpha$ and $\nu_b$ stand for 
 the spectral slope and the break frequency, respectively. This, so-called damped random walk ($\alpha=2$) is widely used to model optical light curves of AGN.
 
\begin{figure}
	\includegraphics[width=\columnwidth]{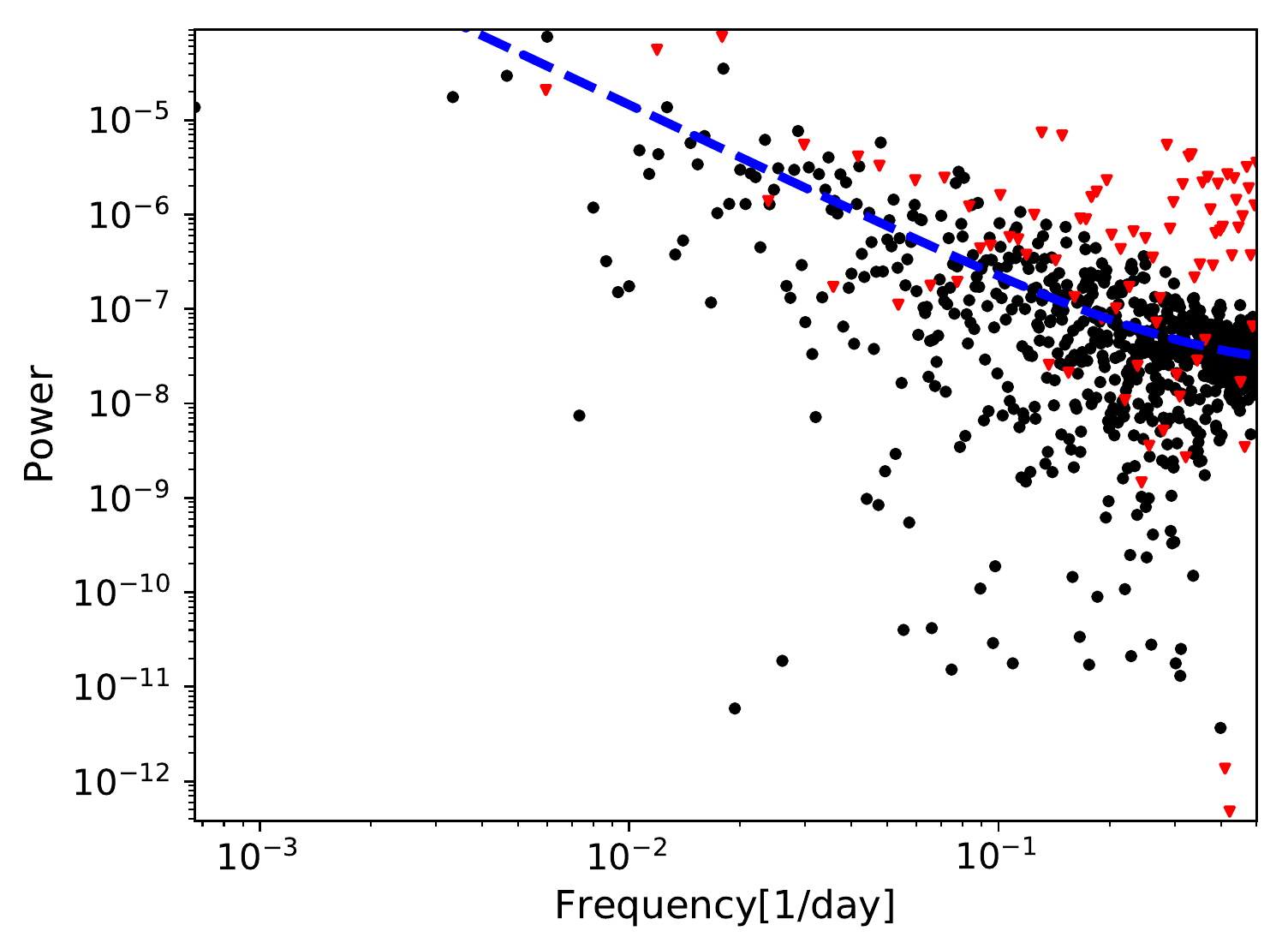}
    \caption{PSD  of artificially upsampled ASAS-SN part of  Mrk 231 light curve (black dots). The best fit model (blue dashed line) has a spectral power-law index 1.85 and break frequency  0.0005 days$^ {-1}$. Red  down triangles mark PSD of original data obtained using the LSP method from the Pyhton module astropy. }
    \label{fig:psdmodel}
\end{figure}

 Fitting was based on the adapted   Basin-hopping of  a stochastic algorithm \citep[see theory in][and for pythonic implementation]{2013MNRAS.433..907E,  ascl.net/1602.012} which attempts to find the global minimum of a given function. As a consistency check, the algorithm can be run from a number of different random starting points so  that the  process of calculation converges to the global minimum.
One thousand  iterations of the Basing-Hopping algorithm is  carried out. Each iteration restarts the minimization from a different value, in order to ensure that the actual minimum is found.
The resulting best-fit
model, corresponding to
$\nu^{-1}_{b}=2000$ days and $\alpha = 1.85$, is shown as the dashed blue  line 
 in   Fig. \ref{fig:psdmodel}. 

For this procedure, we upsampled the  original  non-uniformly sampled data  by fitting   Gaussian process  (GP) with  Ornstein Uhlenbeck kernel \citep{ 2018MNRAS.475.2051K} aka Damped random walk (DRW). This kernel has been widely used in quasar light curve modeling \citep{ 2009ApJ...698..895K, 2010ApJ...721.1014M,  2010ApJ...708..927K}. For seeing the effects of this upsampling, we calculate power spectral density function (PSD) of the original data using the LSP method in the  astropy module of Python (see Fig. \ref{fig:psdmodel}).  There are two dozens of original data points (out of 200)  that are  in subregion parallel to the PSD obtained from GP upsampled curve. The consequence of this fitting and interpolation procedure is that in the same bandwidth as for the original data,  the interpolated data set has a  decreased dispersion.

The  best-fit PSD was used
to generate 1000 artificial  light curves of  Mrk 231, which
has the same sampling, standard deviation and mean as  the observed light curve.
We analyse obtained distribution of this artificial light curves
to infer the significance of the peaks seen in the observed
LSP. We calculate  $95 \%$ and 99.7$\%$ (or $\sim 3\sigma$) statistical significance curve
as the 95th and  99.7th percentile of distribution
of the power in the LSP (see Fig. \ref{fig:LSP}  of the simulated light curves.

Similarly, we calculate the significance of the
observed correlation intensity  of the light curve using the 2DHybrid  method
applied on the 1,000 simulated light curves
from the previously described  PSD model, and their
distribution of correlation intensity  calculated 
in the period-period phase  space (alternatively it can be calculated in frequency-frequency space). The  observed correlation intensity  was compared with  $99.7\%$ confidence level (as the 99.7th percentile at a given period) from the distribution
of the averaged correlation intensity  for the simulated
light curves given in  Fig. (\ref{fig:hyb}). The
significance of the peak centered  at 403 days is above 99.7$\%$.
This approach of  peaks significance determination  is similar to that proposed by \cite{ 2005A&A...431..391V}  for evenly sampled data, but modified here  by using the LSP and 2DHybrid method  instead of the Fourier transform.

\subsection{Analysis of CRTS and ASAS-SN parts of the light curve}

Previous analyses in Subsections \ref{metlsp}-\ref{methyb} were  carried out on the whole data set.
Here we examined CRTS and ASAS-SN data sets separately in the same manner as in the case for the combined light curve.

\begin{figure}
	\includegraphics[width=\columnwidth]{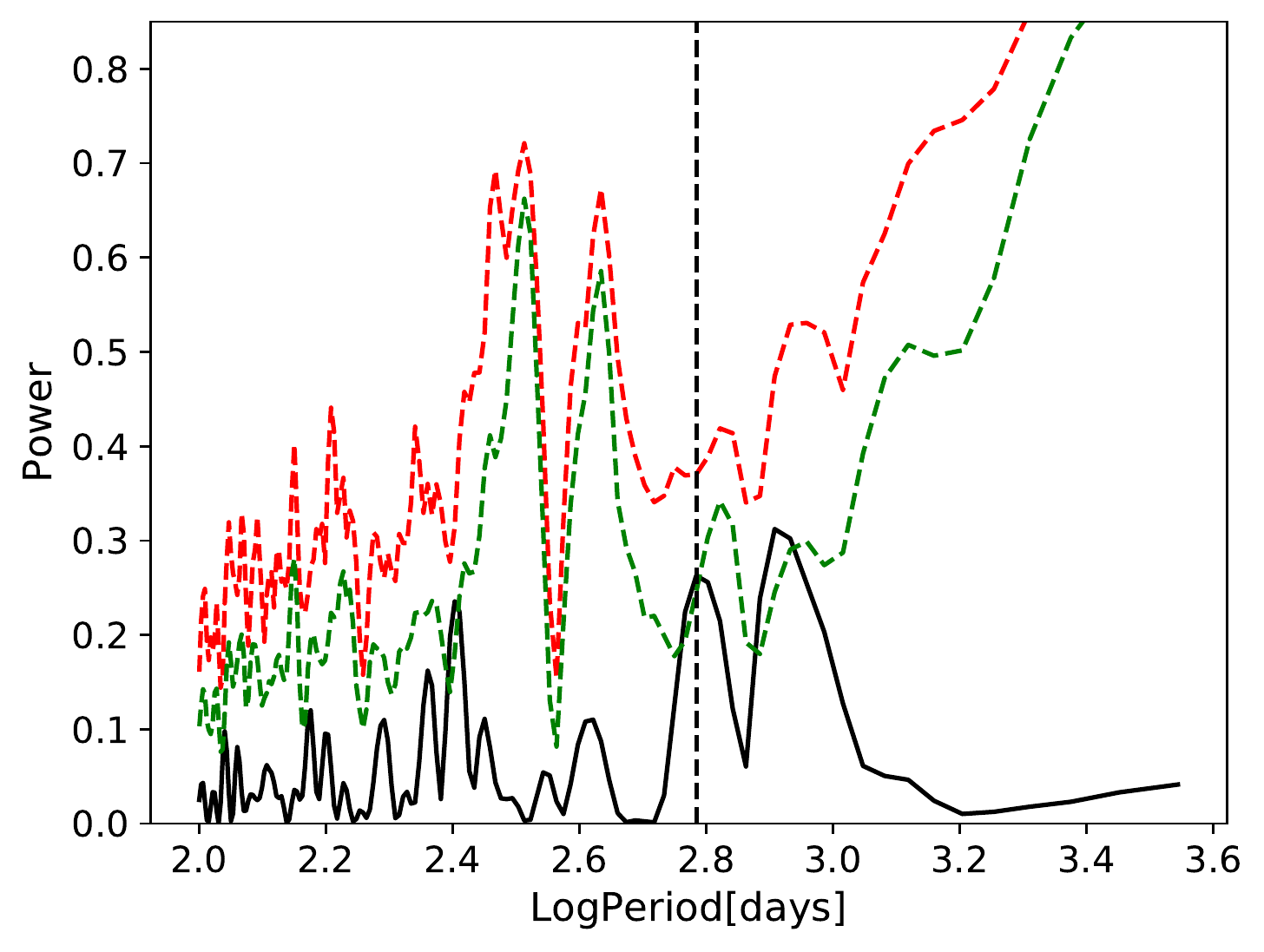}\\
	\includegraphics[width=\columnwidth]{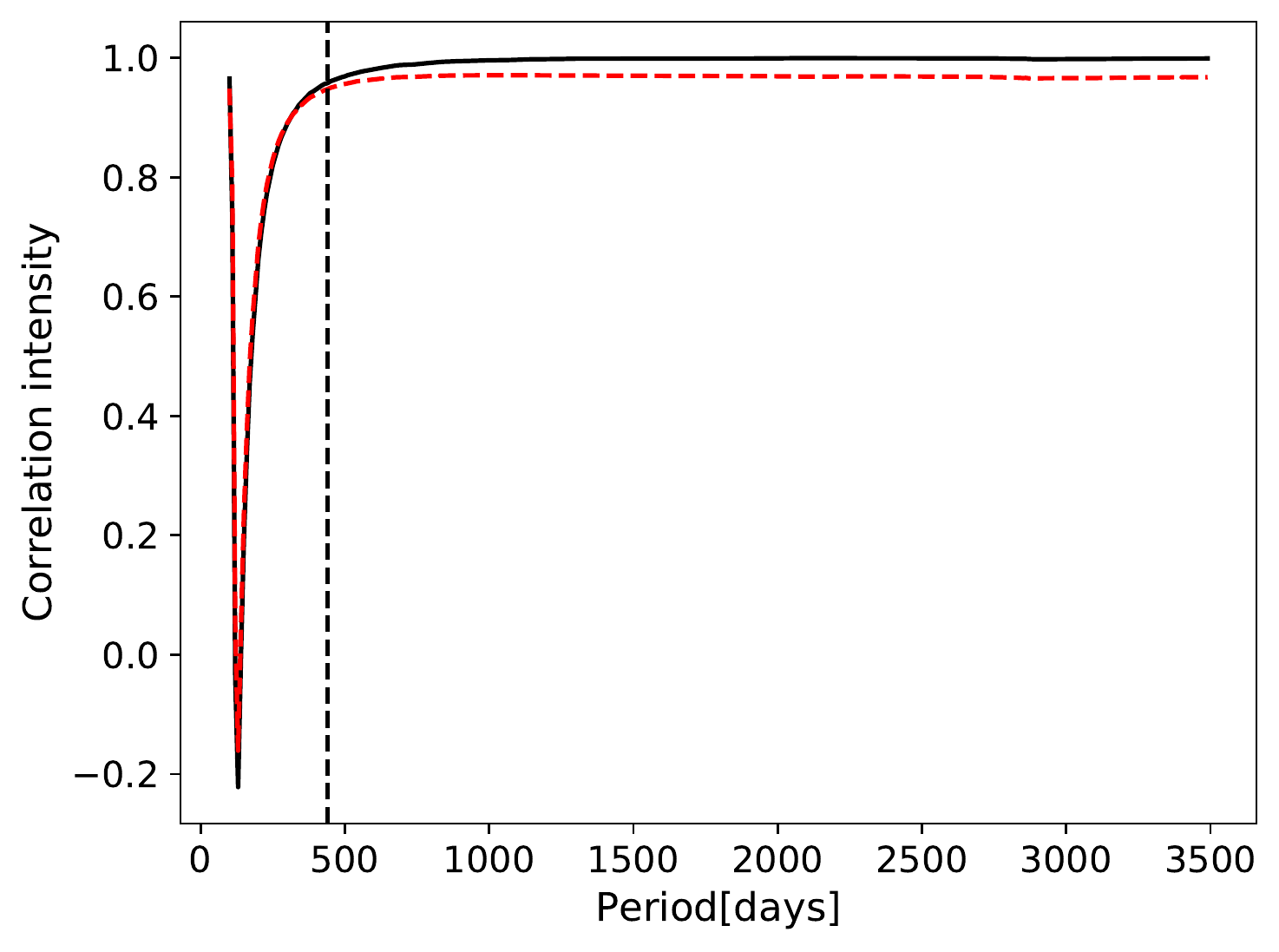}\\
    \caption{Periodicity analysis of CRTS part of the light curve. Upper panel: LSP analysis, the vertical dashed line indicates a period of 609 days. The green (lower dashed) and red (upper dashed)  curve indicate $95\%$ and   $99.7\%$ significance, while the solid black curve is the measured curve.   Lower panel: Hybrid analysis, the vertical dashed line indicates a period of 441 days.  The red dashed curve indicates  $99.7\%$ significance, while the black solid curve is the measured curve.  }
    \label{fig:crts}
\end{figure}

\begin{figure}
	\includegraphics[width=\columnwidth]{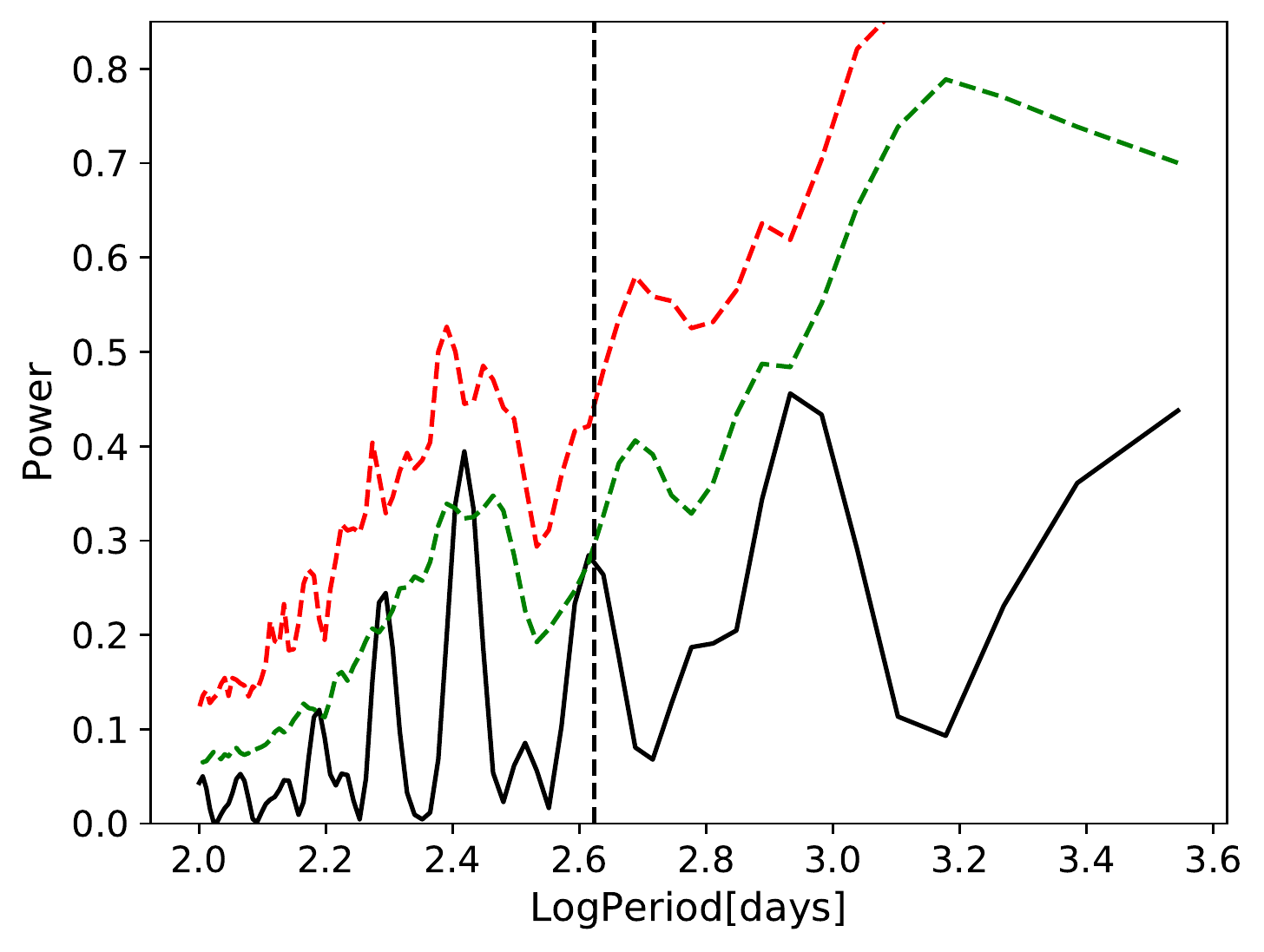}\\
	\includegraphics[width=\columnwidth]{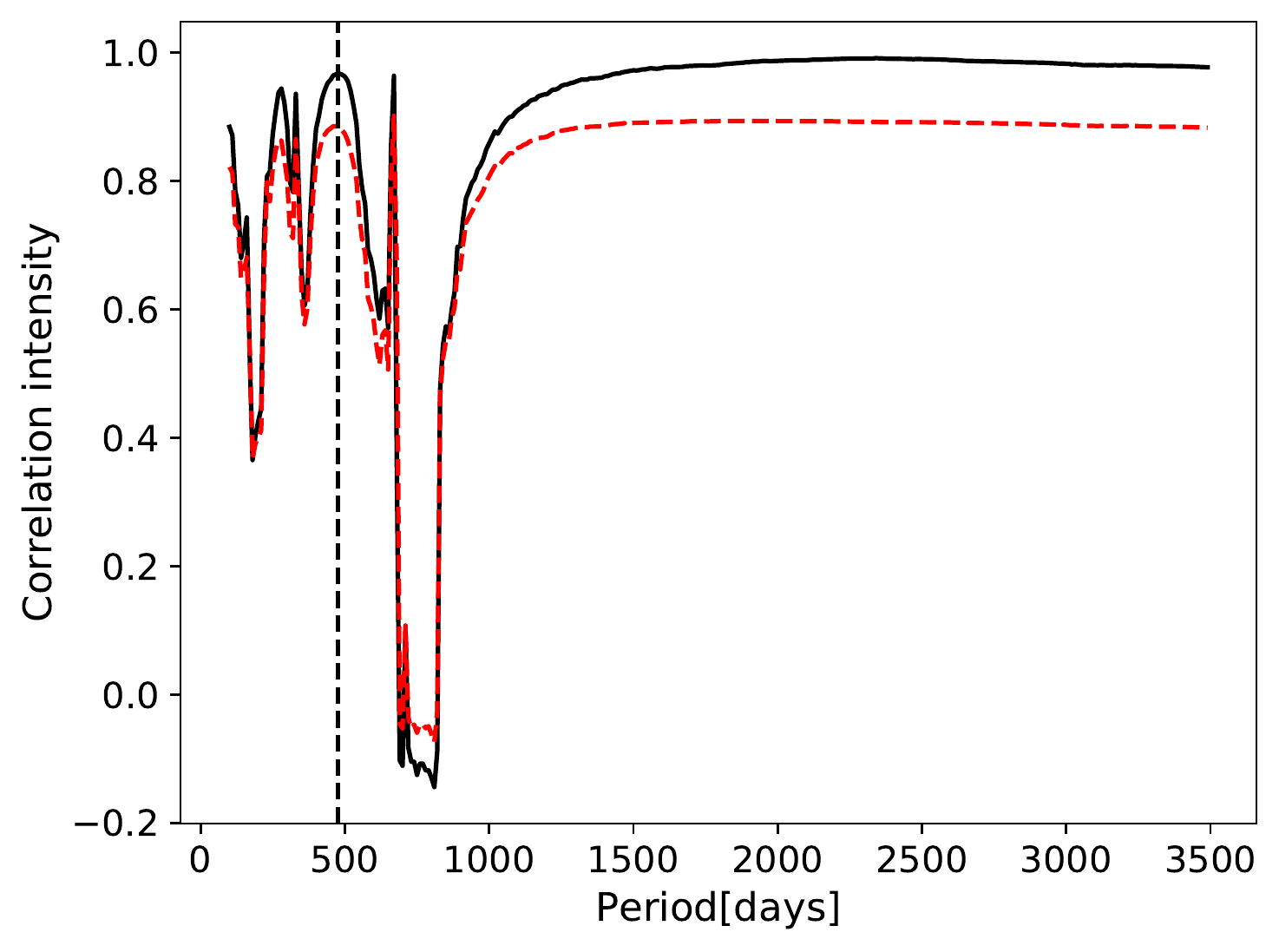}\\
    \caption{Periodicity analysis of the ASAS-SN part of the light curve. Upper panel: the LSP analysis, the vertical dashed line indicates a period of 420 days. The green ( lower dashed)  and red  (upper dashed)  curve indicate $95\%$ and    $99.7\%$ significance, while the  black solid curve is the measured  curve. Bottom panel: 2DHybrid analysis, the vertical dashed line indicates a period of 479 days. The red dashed  curve indicates   $99.7\%$ significance, while the black solid curve is the measured curve. }
    \label{fig:asassn}
\end{figure}

The CRTS part of the light curve shows a period of 609 days in LSP analysis which is above 95$\%$ significance but lower than 99.7$\%$ (see upper panel Fig. \ref{fig:crts}). The 2DHybrid method indicates the presence of a  signal of 441 days above 99.7$\%$ (see bottom panel  Fig. \ref{fig:crts}).
The ASAS-SN part of the curve exhibits the  period of 420 days from LSP analysis, which is barely above $95\%$ but lower than $99.7\%$ (see upper panel Fig. \ref{fig:asassn}). However, the 2DHybrid method detected a strong signal at 479 days, above $99.7\%$ significance.
Based on these results, the ASAS-SN observations have a large influence on the LSP, which is expected due to the smaller uncertainties.
The confidence contours of the 2DHybird method in certain parts  are almost indistinguishable from the correlation intensity of the observed curve in the scales of corresponding figures, as their values are rather close.

\section{Discussion and conclusions}{\label{sum}}
 
For mass ratios $\mathrm{q} >
0.05$, hydrodynamical
simulations indicate that  commensurability patterns  should be present among oscillations in the light curves of binaries \citep{2015Natur.525..351D}.
 The periodogram peaks can  appear as commensurability of  orbital periods of components  $1/2$, as well as in  commensurability of  hotspot period, orbital period and half of the orbital period of the binary \citep[from $ 1/3/6$ to  $ 1/8/16$ see in][]{2015MNRAS.454L..21C}.
In our calcualtions,  both  LSP and 2DHybrid methods  detected three peaks in analysis of combined curve. The peaks at 195 and 403 days are roughly commensurable. The third peak of 265 days is not commensurable with the other two, but it is in rough $3/5$ commensurability with peak of 403 days.
This peak cannot be a consequence SMBBH orbital dynamics; however, it can arise due to some other processes either in circumbinary disk or in disks of components.
\cite{2015ApJ...809..117Y}  interpreted  the UV spectral  signature of Mrk 231 as
a consequence of emission from a circumbinary disc with a central cavity, dynamically opened by  the secondary SMBH. But for 
low mass ratios as it is  assumed to be in the case of Mrk 231, a cavity may not be present as suggested by 
\cite{2016MNRAS.459.2379D, 2015MNRAS.446L..36F}. 
 \cite{10.3847/0004-637X/829/1/}  suggested that the UV emission of Mrk 231 is
more likely to be  consistent with a reddened AGN with a special extinction law. 
Under this model, the UV emission from the nucleus is completely obscured by the
dust and the observed UV flux is interpreted from starburst activities from the host
galaxy, which does not forecast variability on short timescales.  \cite{2018MNRAS.480.5504Y}
analyzed Swift data and detected a significant UV variability of Mrk 231 consistent
with accretion disk emission with a characteristic red-noise spectrum.  Both, our
result  and \cite{2015ApJ...809..117Y}  SMBBH model  suggest that  a possible 1.1 $\sim$1.2 yr  periodic
variation   could be also detected in the UV
emission of Mrk 231; however,  probing this frequency is beyond the range
of  current UV data \citep[see][]{2018MNRAS.480.5504Y}. The detailed UV emission location is
still uncertain, although it is largely constrained within the central engine, and
results of reverberation mapping campaigns can help to identify AGN
variability model. For instance, any indication of the UV/optical
variations lagging behind  the X-ray ones in Mrk 231 emission would
support scenario of variability due to  reprocessed X-ray emission by the
accretion disc \citep[see][]{2018MNRAS.480.5504Y}.
The signal of 1.1 yr is always present in our  LSP analysis  with an almost unaltered  amplitude.
Previous studies of  red noise light curves show that larger LSP peaks are more likely to appear at low frequencies \citep{2008ApJ...688L..17M}, but we do not observe this phenomenon in   Mrk 231 light curve. Also, astronomical periodic signals often 
significantly depart from sinusoids. Thus, the results of the LS periodogram
is less optimal compared to 2DHybrid method, since the periodic variation would
be fitted inadequately. This is seen in  the  light curves
of some  variable stars  and in the radial velocity curves
of stars orbited by a planet on an eccentric orbit \citep[see details in][]{10.1111/j.1365-2966.2009.14634.x}. Also, \cite{10.1111/j.1365-2966.2005.09197.x} showed that a simple damped oscillator represents well the time-resolved light-curve properties of Cyg-X1.
We use the complex  Morlet wavelet, which means that we probe the signal which is deformed in the 2DHybrid analysis.

If the detected  periodicity ($\mathrm{P}$)  is  assumed as the redshifted (z) orbital period of a  SMBBH  orbital period ($\mathcal{O}$) \citep{10.1038/nature14143}, we can estimate some parameters related to the hypothetical binary system in Mrk 231. As listed in Table \ref{tab:SMBBHparam}.  we give the   orbital period $\mathcal{O}={P}/{(1+z)}$; SMBBH mutual distance  if the motion is circular
\begin{equation}
 \mathrm{r}=\left(\frac{GMP^{2}}{4 \pi^{2}}\right)^{\frac{1}{3}},
\end{equation} 
\noindent where G and M are the gravitational constant and  the total mass of the assumed binary system.
 We also calculate  the
maximum angular separation  if  a circular orbit is assumed
  $\theta={\mathrm{r}}/{\mathrm{D}_{\mathrm{A}}}$, where  $\mathrm{D}_{\mathrm{A}}$  stands for  the angular size distance in  the standard cosmology of a flat universe having Hubble constant $H_{0}=67.8\, \mathrm{km }\,\mathrm{s}^{-1} \mathrm{Mpc}^{-1}$ and a matter density parameter $\Omega_{m}=0.308$ \citep{10.1051/0004-6361/201525830}.
Finally, we calculate 
the time of coalescence of  SMBBH, namely the time  at which, formally, the relative distance  due to gravitational waves emission  vanishes \citep{10.1103/PhysRev.136.B1224}:
\begin{equation}
\tau_{GW}=\frac{5}{256}\frac{c^3}{G^3}\frac{\mathrm{r}^4}{(M_{1}+M_{2}) M_{1}M_{2}},
\end{equation}
where $M_{1}$ and $M_{2}$ are masses of the primary and secondary component, respectively.

\noindent \cite{2016MNRAS.463.2145C} identified a statistically significant population of 50 periodically variable quasars.
These objects show   short periods of a few hundred days. Also, they found that  the  distribution of periods 
of this population supports SMBBHs with a low
mass ratio ($q \sim 0.01$). Although hydrodynamical simulations of SMBBHs with larger  mass ratios $q > 0.3$  \citep[e.g.][]{ 2012A&A...545A.127R,2014ApJ...783..134F}  predict that   the optical periodicity may not always arise due to  the binary orbital
period but it can be a consequence of some other superimposing  phenomena in binary systems (e.g. lump in the lopsided accretion disc). In general,
periodic variability of quasars can also be explained as   quasi-periodic modulations
 arising from surrounding  regions of a single SMBH due to  Lense-Thiring precession,
a warped accretion disc, or the precession of a jet
\citep{2015MNRAS.453.1562G}.

Interpreting our results in combination with those from other studies of  highly unequal -mass SMBBH systems  \citep{2012MNRAS.427...77V, 2015Natur.525..351D,2016ApJ...827...56Z,2015Natur.525..351D},  invokes the highly unexpected scenario  that
SMBBHs with low mass ratios may be more frequent  than equal mass ratio systems.
Our  results are also  a consequence of  the better photometry  and
the better sampling of the ASAS-SN data compared
to the CRTS.
In this work, we identified a period of  403 days  in the optical light curve of Mrk231. 
This period  obtained from the timing analysis matches  \cite{2015ApJ...809..117Y} prediction, which strengthens the SMBBH hypothesis.

However, due to  the
limited photometric accuracy of the data from
CRTS, it seems  crucial to further follow this object to confirm its periodicity.
Also, in other bands such as UV, \cite{2018MNRAS.480.5504Y} have detected UV variability of Mrk 231. An independent UV periodicity confirmation will further validate the SMBBH nature of Mrk 231.

\begin{table*}
	\centering
	\caption{Summary of light curve characteristics and detected periods. L,  $\mathrm{mag}_{mean}$, $\mathrm{\sigma}_{mean}$ are the total time baseline, mean magnitude and mean  magnitude error of the light curves, respectively; N,  $\mathrm{s}_{mean}$  and  $\mathrm{s}_{median}$  are the number of data points, mean and median sampling of the light curves; $\mathrm{F}_{var}$  is the fractional variability of the light curves;  the last four columns provide results  for period detection based on LSP and 2DHybrid methods  including the significance of the peaks (sig), respectively. }
	\label{tab:prva}
	\begin{tabular}{cccccccccccc} 
		\hline
		Data & L&$\mathrm{mag}_{mean}$&$\mathrm{\sigma}_{mean}$&N&$\mathrm{s}_{mean}$&$\mathrm{s}_{median}$&$\mathrm{F}_{var}$&LSP &sig$_\mathrm{LSP}$ &2DHybrid &sig$_\mathrm{2DHyb}$\\
		 & [days]&&&& [days]& [days]&&[days]&[$\%$]&[days]&[$\%$] \\
		\hline
		CRTS+ASAS-SN &4782 &13.25&0.03&277&17.33&2.98&0.0023& 416$\pm $34 &95& 403$\pm$29&99.7 \\
				CRTS & 2947&12.95&0.05&113&26&0.006& 0.0011& 609$\pm$ 54 & 95&441$\pm $201&99.7\\
		ASAS-SN &1587&13.26&0.017&164&9.5&3.98& 0.0029&413$\pm$ 40 &95 &475$\pm$ 75&99.7 \\

		\hline
	\end{tabular}
\end{table*}

\begin{table*}
	\centering
	\caption{SMBBH properties based on a  detected period of $\sim$403 days. Columns are described in the text. Masses  for primary $M_1$ and secondary $M_2$ components are taken from \citet{2015ApJ...809..117Y}.} 

\label{tab:SMBBHparam}
	\begin{tabular}{cccccc} 
		\hline
		$\mathrm{M}_{1} $ & 	$\mathrm{M}_{2}$  & $\mathcal{O}$&$\mathrm{r}$&$\theta$&$\tau_{GW}$ \\
		 $\left[\mathrm{M}\odot \right]$&  $\left[ \mathrm{M}\odot \right]$ & [days]&[pc]&$[\mu as]$& [yr]\\
		\hline
		$1.5\cdot 10^8$  &$4.6\cdot 10^6$ & $\sim$403& 0.00294&3.5&$4.1\times10^5$ \\
		\hline
	\end{tabular}
\end{table*}

\section*{Acknowledgements}

This work is supported by
Science and Technological development of Republic Serbia  through the  
\emph{Astrophysical Spectroscopy of Extragalactic Objects} project number 176001, and by  the National Natural Science Foundation of China (grants 11863007) and  the Ministry of Education.
XD acknowledges the financial support from NSF grant AST-1413056.
ASAS-SN is supported by the Gordon and Betty Moore
Foundation through grant GBMF5490 to the Ohio State
University, and NSF grants AST-1515927 and AST-1908570. Development of
ASAS-SN has been supported by NSF grant AST-0908816,
the Mt. Cuba Astronomical Foundation, the Center for Cosmology
and AstroParticle Physics at the Ohio State University,
the Chinese Academy of Sciences South America Center
for Astronomy (CAS- SACA), the Villum Foundation, and
George Skestos.









\bsp	
\label{lastpage}
\end{document}